\documentclass[11pt,a4paper,reqno]{amsart}
\usepackage[hmargin={2.7cm,2.7cm},vmargin={2.5cm,2.5cm},centering]{geometry}
\usepackage{amssymb,amsmath,amsthm,amsfonts,amscd}

\let\a=\alpha \let\b=\beta  \let\g=\gamma  \let\d=\delta
\let\e=\varepsilon
  \let\h=\eta     \let\l=\lambda
\let\m=\mu    \let\n=\nu    \let\x=\xi         
\let\s=\sigma     
   
 \let\D=\Delta  \let\L=\Lambda

\def\EE{{\cal E}}

\def\EE{\mathbb{E}}
\def\\{\hfill\break}
\def\={:=}
\let\io=\infty
\let\0=\noindent

\def\tende#1{\,\vtop{\ialign{##\crcr\rightarrowfill\crcr\noalign{\kern-1pt
    \nointerlineskip} \hskip3.pt${\scriptstyle #1}$\hskip3.pt\crcr}}\,}
\def\otto{\,{\kern-1.truept\leftarrow\kern-5.truept\to\kern-1.truept}\,}

\def\to{\rightarrow}

\def\qed{\hfill\raise1pt\hbox{\vrule height5pt width5pt depth0pt}}

\def\be{\begin{equation}}
\def\ee{\end{equation}}
\def\bp{\begin{pmatrix}}
\def\ep{\end{pmatrix}}
\def\bea{\begin{eqnarray}}
\def\eea{\end{eqnarray}}

\def\pref#1{(\ref{#1})}

\usepackage{latexsym}
\usepackage[english]{babel}
\usepackage{fancyhdr}
\usepackage[mathscr]{eucal}
\usepackage{amsmath}
\usepackage{mathrsfs}
\usepackage{amsthm}
\usepackage{amsfonts}
\usepackage{amssymb}
\usepackage{amscd}
\usepackage{bbm}
\usepackage{graphicx}
\usepackage{graphics}
\usepackage{latexsym}
\usepackage{color}
\usepackage{pifont}
\usepackage{color}
\usepackage{tikz}
\usetikzlibrary{patterns, decorations.markings,decorations.pathmorphing,arrows, calc}

\tikzset{snake it/.style={decorate, decoration=snake}}

\newcommand{\ZZ}{\mathbb{Z}}

\theoremstyle{plain}

\theoremstyle{definition}

\newtheorem*{remark*}{Remark}

\numberwithin{equation}{section}

\usepackage{amsmath,amsthm,amsfonts,amssymb, graphicx,mathabx,epsfig}
\usepackage{bbm}
\usepackage{bm}
\usepackage{graphicx,color}
\usepackage{footmisc}
\usepackage{epstopdf}
\usepackage{dsfont}
\usepackage{pgfplots}
\usepackage{subfig}
\usetikzlibrary{fit}
\usepackage{slashed}
\usepackage{amsaddr}

\makeatletter
\renewcommand{\email}[2][]{%
  \ifx\emails\@empty\relax\else{\g@addto@macro\emails{,\space}}\fi%
  \@ifnotempty{#1}{\g@addto@macro\emails{\textrm{(#1)}\space}}%
  \g@addto@macro\emails{#2}%
}
\makeatother

\title{Renormalization: general theory}

\author{Vieri Mastropietro}
\address{Dipartimento di Matematica F. Enriques - 
Universit\`a di Milano \\ via C. Saldini 50 - 20129 Milano (Italy).}
\email{vieri.mastropietro@unimi.it}

\date{\today}

\begin{document}

\maketitle

\vspace{-0.5cm}

\begin{abstract}
We review the theory of renormalization,
including perturbative renormalization, regularized
functional integrals, Renormalization Group and rigorous renormalization.
\end{abstract}

\tableofcontents

\section{Introduction}

The notion of {\it Renormalization} was introduced  originally
by Bethe, Schwinger, Feynman, Tomonaga, Dyson 
and others in the 1950's (see e.g. \cite{B}, \cite{D1}), in the context of Quantum Electrodynamics (QED)
and it was crucial for particle physics in the 1960s and 1970s.
This concept had
crucial developments in the 1970s with the introduction of the 
{\it Renormalization Group} by Kadanoff \cite{Ka} and 
Wilson \cite{W} and the applications to statistical physics.
Starting from the late 1980s it has been applied to Condensed Matter and Mathematical Physics.

A common feature of problems where Renormalization appears
is the presence of infinitely many particles (real or virtual)
in interaction between them, often with an 
emergent behavior which can be very different at different energy scales.
In particle physics Renormalization is essential for the formulation of the Standard Model,
the theory describing the interaction of all the known elementary particles
through
the Electromagnetic, Weak and Strong forces, that is all known forces except gravity. In statistical physics it is at the basis
of the theory of classical phase transitions and critical phenomena and the corresponding universality properties.  In condensed matter, it has been applied to the understanding of quantum liquids or
phases with broken symmetries like superconducting states. In mathematical physics, it is one of the
very few techniques at the basis of the rigorous construction of theories of modern physics.
More in general, 
Renormalization and Renormalization Group have provided 
an unifying
language for phenomena apparently very different and have given
a method to link physical descriptions at different scales, 
providing a quantitative foundations to 
concepts like
universality classes, relevant or irrelevant interactions
and emergent behavior.

\section{Perturbative renormalization}

QED is the Quantum Field Theory (QFT) describing the interaction of
electrons (or more in general fermionic particles) and electromagnetic (e.m.) fields, combining the principles of quantum mechanics with special relativity in a 4 dimensional space-time. Both particles and e.m. fields are described in terms of quantized field operators and the excitations of the e.m. one are the photons.
The theory is described by an action (called bare action)
given by sum of monomials in the fields and depending on
a number of parameters $v_b=(e_b, m_b, Z^\psi_b, Z^A_b)$, representing respectively the electron 
charge and  mass, and the electron or photon
wave function normalization. The form of the bare interaction is dictated
by the request of Gauge invariance based on $U(1)$ symmetry, which
is a crucial feature of classical electromagnetism.

In absence of interaction between electrons and photons (free fields) the physical
properties, denoted by $P$, can be computed in an exact way, but  this is not possible in presence of interaction.
The charge, which provides the
coupling between the electron and photon, is rather small in adimensional units
and this suggest to include interactions
by a perturbative method, 
writing the physical observables as a Taylor series 
in the charge 
\be
P=P_0+ e_b P_1+e_b^2 P_2+...\quad\quad P_n\equiv P_n(m_b, Z^\psi_b, Z^A_d)
\ee
Each term $P_n$ can be written, according
to precise rules, as sum of integrals over a set of energy-momentum variables $k_i\in \mathbb{R}^4$, where $i$ is an index assuming $O(n)$ values,
of a 
product of functions ("propagators") $\hat g(k)$ with $k$ equal to one of the momenta or a linear combination of them.
Each of such integrals can be conveniently also represented in a graphical way ("Feynman graphs").
The propagators $\hat g(k)$ vanish as $k\to \io$ as the inverse of the momentum 
with power one or two depending if the propagator is associated to electrons or photons. This slow decay has the effect that the integrals
contributing to $P_n$ are often diverging with an infinite value.
This divergence is called
"ultraviolet" problem (as related to the large energies); there is also a divergence associated to the singularity
of $\hat g(k)$ for massless particles at $k=0$ , which is called 
"infrared" problem. 

The presence of divergences makes the outcome of the 
theory apparently useless, and the renormalization procedure was invented to extract meaningful
results. The first step consists in 
"regularizing" the integrals contributing to the observable $P$, 
introducing a parameter $\L$,  
called ultraviolet cut-off, making the integrals 
finite, at least if all the particles are massive; the divergence is recovered
in the limit in which $\L$ is sent to infinity (one says "removing the cut-off"). There are several ways in which a cut-off can be introduced; typically certain regularizations break some symmetries formally present without cut-offs and respect others.
The simplest regularization consists simply in replacing $\hat g(k)$ with $\chi(k)\hat g(k)$,
where $\chi(k)$ vanishes for $|k|\ge \L$ (momentum regularization); another regularization
consists in varying the dimension of the integrals to make them finite (dimensional regularization) or replace the continuum space-time with a discrete one
(lattice regularization).

Physical quantities become therefore functions 
of the parameters $v_b$ and the cut-off $\L$, that is 
$P=P(v_b,\L)$. The interaction modifies the mass, the  charge and and the normalizations; such quantities are indeed among the physical observables $P(v_b,\L)$
one can compute. They are called  dressed or renormalized values $v_d$
(in contrast to the free or bare ones $v_b$), 
and expressed by
series expansion which are functions of $v_b$ and $\L$, that is 
$v_d=v_d(v_b,\L)=v_0+e_b v_1+e_b^2 v_2+..$ with $v_n(m_b, Z^\psi_b, Z^A_b)$. 
The crucial observation is that the parameters which are measured in the laboratory are the dressed quantities $v_d$, not the bare ones $v_b$;
the latter are ideal objects corresponding
to the unphysical situation
in which the interaction of charged particles with the e.m. field can be switched off. 
One can therefore invert order by order the relation $v_d=v_d(v_b,\L)$ 
expressing the bare parameters in terms of the dressed ones and the cut-offs
$
v_b=v_b(v_d,\L)$: finally
we can insert then this expression in the observables obtaining
\be
P(v_b(v_d,\L),\L)\equiv \bar P(v_d,\L)=e_d \bar P_1
+e_d^2 \bar P_2+...\quad \quad \bar P_n=\bar P_n(m_d, Z_d^\psi, Z^A_d,\L)\label{sss1}
\ee
that is the observables are expressed in term of the dressed couplings
$e_d, m_d,  Z_d^\psi, Z_d^A$.
The remarkable property is that $\bar P_n$ is {\it finite}
removing the cut-off $\L\to\io$; that is all the divergences of any observable can be reabsorbed in a redefinition of the 
parameters of the theory. If one fixes $e_d, m_d$
equal to measured charge and mass of the electron (denoted simply by $e$ and $m$) and 
$Z_d^A=Z_d^\psi=1$ then it is possible to 
get predictions
which can be experimentally tested, and the results are in impressive
agreement with experiments.
A celebrated example is the electron magnetic moment, in which the experimental deviation with the respect to the prediction of the non-interacting theory was explained by QED with the renormalization procedure with amazing precision.

An important feature of QED is that the bare parameters are not independent one from the other but are connected by non-trivial relations; in particular
\be
e_b=e_d/\sqrt{Z^A_b} \label{imp}
\ee
that is the bare charge is the dressed one divided by the bare photon wave function renormalization.
The above relation is essential for physical consistence. There are several elementary particles in nature, and it
turns out the the ratios of the charge is a rational number
(quantization of the charge), so that by \pref{imp} this is true also for bare charges.
Such relation is a consquence of special relations following from Gauge invariance and known as Ward Identities.

The property that the observables can be made order by order finite
with a proper choice of the bare parameters is called 
{\it perturbative renormalizability}. One can check explicitly that this is the case
at lowest order in perturbation theory and a general proof for QED
was given in \cite{B1}. A general criterion for the perturbative renormalizability of a QFT
is based on the superficial degree of divergence D of a Feynman graph, which is simply the number of integrated momenta minus
the total decay power in momenta due to the propagators.
The criterion says that theory is renormalizable if $D$ is independent
from the perturbative order $n$ and is positive or vanishing only for 
Feynman graphs with a number of external lines corresponding to monomial in the fields as in the bare action.
A basic example of QFT satisfying this criterion is 
the $\phi^4$ theory in 4 dimension, describing the self-interaction of a scalar boson field, where $D=4-n_A$, if $n_A$ is the number
of external boson fields; $D=0$ for $n_A=4$, $D=2$ for $n_A=2$ corresponding to terms associated to bare parameters (the bare coupling and mass), while $D<0$ for $n_A\ge 6$. 

The criterion for renormalizability 
is rather simple and intuitive, but the proof that the verification of the condition implies the perturbative renormalizability
relies on subtle combinatorics \cite{H1},
\cite{Z}. The divergence in the graphs are due to divergences in subgraphs
encapsulated one into the other which must be compensated by suitable terms
appearing in the bare parameters at any order.
The combinatorics was greatly simplied
using the Wilson RG approach (see below) in \cite{G} using the tree expansion 
and in \cite{P} using flow equations. 
Such methods were applied to QED respectively in  
\cite{H} and \cite{K}. In contrast to $\phi^4$, however,
the simple criterion of superficial divergence cannot be applied;
in QED one has $D=4-3 n_\psi/2-n_A$, with $n_\psi, n_A$ the external lines corresponding 
to electrons or photons: it is
$n$-independent but $D\ge 0$ also in correspondence of terms not
in the action, like $n_A=4,2$ and $n_\psi=0$. Using dimensional regularization
the degree of divergence can be improved and one can show that such terms are indeed vanishing,
as an effect of gauge invariance, together with \pref{imp}. There is however
a basic tension between
the momentum decomposition typically used in RG and gauge invariance, and this requires 
introducing extra parameters breaking  Gauge invariance in the bare action: Gauge symmetry
is recovered at the level of
identities (Ward Identities) between the physical observables only when the ultraviolet cut-off is removed.

Not all QFT are perturbatively renormalizable, and indeed only 
a small number of QFT has this property.
A well known example is the Fermi theory
\cite{F} 
describing the 
nuclear Weak forces, appearing for instance in the phenomenon of beta decay in nuclei.
In contrast with
e.m. forces, they are short ranged and are described in the Fermi theory 
by a purely fermionic local quartic interaction. In this case
the superficial degree of divergence is $D=4+2 n-3 n_\psi/2$, that is it increases with the order: the theory is non-renormalizable and one cannot absorb the divergences in the bare parameters.
The physical observables can be written as series
of the Fermi coupling $G$ and are diverging with the ultraviolet cut-off $\L$, that is
\be 
P(G,\L)=\sum_{n=1}^\io 
 G^n P_n \quad\quad P_n=O(\L^{2 (n-1)})
\ee
As the lowest order of several observables is independent on the cut-off, the
theory can be anyway used 
up to a certain energy scale, naturally determined by the condition $G \L^2<1$. 
A theory valid up to a certain energy (or length) scale is called 
{\it effective}.  
The Fermi theory is perfectly valid in this regime of validity, which was suitable for the energies reached in experiments in the 1960s and had
crucial applications in particle physics \cite{GIM}.

It was natural, inspired by QED, to look for a theory for weak forces with the additional requirement of Gauge invariance and
renormalizability. The symmetry is dictated by the form
of currents involved in weak forces, and a good candidate for describing
the e.m. and weak forces is the Yang-Mills gauge theory based on the 
non abelian $U(1)\times SU(2)$ gauge symmetry. This leads to the introduction, in addition to photons,
of other gauge bosons named $W,Z$ corresponding to the interaction with weak forces.  However
the symmetry which dictates the action of such theory
forbids the presence of mass terms in the action; the mass of the Gauge bosons breaks the Gauge symmetry, and the mass of the fermions breaks the chiral symmetry, related to the fact that the interaction with left or right handed particles is different. Such masses are however experimentally observed. In 
the Standard Electroweak theory \cite{We} of Weinberg and Salam 
there are no mass term in the bare action but there is an 
extra field, the Higgs field, described by a scalar self-interacting theory similar to $\phi^4$,
such that the masses are generated by the interaction with it without breaking the symmetry in the bare action. The theoretical mechanism how this is achieved is the one of the spontaneous symmetry breaking, at the basis 
of the theory of phase transitions in statistical mechanics.

The symmetry principles are then verified and gauge invariance holds, 
but 
the theory is however still {\it not} perturbatively renormalizable
unless  the parameters appearing in it verify certain relations. The reason why this is so can be understood by considering a Gauge theory with massless fermions and a non vanishing boson mass.
Let us start from a non chiral theory like QED: if
a mass to the photon is added then
gauge invariance is lost and 
the (Euclidean) propagator have the form
\be
{1\over k^2+M^2}(\d_{\m\n}+ {k_\m k_\n\over M^2} )
\ee
where $k_\m$ are the components of $k$ and $M$ is the mass.
Due to the presence of the second term, the propagator does not vanish as $k\to \io$, in contrast with the massless case in which it decays as
$O(k^{-2})$. The degree of divergence $D$ becomes the same as the one in Fermi theory
and one gets again a non-renormalizable theory. However
if the fields in the bare action would be classical then by symmetry
the current is conserved $k_\m \hat j_\m=0$;  at a quantum level
this conservation appear in the form of Ward identities implying that,
if one restricts to gauge invariant observables ,
then
the contribution from the second term in the propagator is vanishing. Renormalizability is therefore preserved in a non chiral gauge theory like QED even in presence of a photon mass. This argument is however
generically non valid 
in a chiral Gauge theory. 
One could add a bare mass to a chiral gauge theory (with vanishing mass for fermions)
and again classically the chiral current is conserved. However at the quantum level such current is {\it not}
conserved, a phenomenon called {\it quantum anomaly}
\cite{Ad}, and such breaking destroys the renormalizability.
The same problems are encountered
if one generated the mass of the bosons via the interaction with the Higgs mechanism in the Standard Electroweak sector.

A peculiar property of the anomaly is that its value is exactly given
by its lower order contribution, that is all higher orders contribution vanishes order by order,
a property known as {\it anomaly non-renormalization} and proved using dimensional regularization. 
 Therefore its value can be exactly computed and it turns out that, if certain condition on 
the charges of the particles is verified, the anomaly cancels out \cite{1} and the theory is again renormalizable.
The condition is 
\be
\sum_i Y_i^L-\sum_i  Y_i^R=0\quad\quad \sum_i (Y_i^L)^3-\sum_i  (Y_i^R)^3=0
\label{appa}
\ee
where $Y_i^L$ and $Y_i^R$ are the hypercharges of the particle of type $i$.
As there is a simple relation between hypercharge and electric charges, this relation implies a relation between charges.
This validity of this relation severely constraints the type of elementary particles in the
theory and the respective charges, and is verified by the 
elementary particles experimentally discovered
which are (three families) of two leptons (for instance neutrino and electrons) and two quarks
(constituing the protons and neutrons), which charges given by, in suitable units, 
$(0,-1, 2/3,-1/3)$. As a consequence
the Standard Electroweak theory is perturbatively renormalizable \cite{TH}, 
using dimensional regularizations. 
In conclusion
the requirement of perturbative renormalizability
leads therefore to the introduction of a number of particles, like the bosons $W,Z$ snd the Higgs, which were observed years after,
and provide a partial explanation to charge quantization. 

The nuclear Strong forces, which glue together particles in the atomic nuclei,
were also described as a gauge theory for quarks associated
to the non-abelian $SU(3)$ symmetry, a  theory called Quantum Chromodynamics (QCD) The fermions are coupled to boson fields known as gluons
and the associated currents are non chiral. Again the mass of the fermions and the gluons
are not present in the bare action, in order to preserve gauge invariance; they are generated via a spontaneous symmetry breaking mechanism, proposed in
\cite{Na} and related to the superconductivity phenomenon in solid state physics. 
The electroweak and QCD theories are therefore renormalizable theories
and form the Standard Model of elementary particles.
Physical
properties are expressed by series order by order finite as $\L\to\io$; by truncating them one can make predictions, neglecting
the contribution of the remaining terms. 

There is however no mathematical guarantee that the remainder 
which is neglected is small or even finite; the series are not convergent and probably even not asymptotic, a property which
would imply some control on the size of the rest.
It was indeed observed in $QED$ and in $\phi^4$ theory that, if a partial resummation of the perturbative series is done,
the bare coupling is diverging at a {\it finite} value of cut-off.
In the case of $\phi^4$ theory, 
it has been indeed proved that 
the only way in which one can remove the ultraviolet cut-off 
is obtaining a theory with vanishing dressed coupling, that is not interacting, see \cite{21},\cite{22}. Therefore at least in the case of $\phi^4$ the series are surely not asymptotic expansions of a QFT.

It is expected that the cut-off can be removed in 
QCD, that is the sector of the Standard Model describing the strong forces;
the bare charge is decreasing with the increasing of the cut-off, a phenomenon called asymptotic freedom
\cite{Na1} and mass is generated via a different mechanism not requiring the interaction with the Higgs.
However
the Standard Model as a whole could be also trivial, hence the cut-off cannot be removed beyond perturbation theory. 

This rises the question of why
the predictions obtaining truncating the series are so in agreement with experiments.
The actual idea is that also the Standard Model has to be considered as an effective theory, valid only with a finite cut-off;
the counterpart of renormalizability is that the cut-off can be chosen much higher than in non-renormalizable series,
so that its effects are beyond any possibility to be observed. 
This seems reasonable also independently from triviality, as at very high energy surely gravity should be taken into account and the QFT could be 
replaced by something else.
This point of view appears naturally in the RG approach described below.

\section{Euclidean functional integrals}

The Renormalization Group approach \cite{W} provides a different interpretation of renormalization beyond 
formal perturbative expansions and a natural setting is given by
functional integrals.
In the case of QFT the starting point is the path integral
formulation of Quantum mechanics. In classical mechanics 
the trajectories of a system are the ones minimizing (or more in general making stationary) a certain action $S=\int dt \mathcal{L}$ where $\mathcal{L}$ is the
Lagrangian of the system. In quantum mechanics 
the probability amplitude is given by summing
all paths weighted by a complex factor $e^{i S \over\hbar}$, and in the case of QFT this lead to an "integral over histories" which is generally not well 
defined. However one can obtain a well defined expression by
passing from to the Minkowski spacetime 
to
an Euclidean metric performing a "Wick rotation", that is replacing 
in the space time vatiable $t \to i x_0$. In addition a lattice with finite step and volume is introduced,
so that one arrives to a set of Euclidean correlations of the following 
form
\be
S(x_1,..,x_n)={\int P(d\Phi) e^{-V(\Phi)} \Phi_{x_1}...
\Phi_{x_n}\over \int P(d\Phi) e^{-V(\Phi)}  }\label{1}
\ee
where $x\in \L$ are discrete space-time coordinates, for instance
$\L=a \ZZ^d\cap [0,L]^d$, with $a$ the lattice step.
The correlations are then obtained 
averaging over all the field configurations with weight $P(d\Phi) e^{-V(\Phi)}$, while the partition function $Z$ is the denominator of \pref{1}.

Two main cases are possible respectively called bosonic or fermionic. 
In the bosonic case ($\Phi=\phi$) the collection of all $\phi$ belongs to
$\mathbb{R}^{|\L|}$, $P(d\phi)$ is a Gaussian measure
and $V(\phi)$ is a sum over monomials in $\phi$. An example is provided by the lattice $\phi^4$, where 
\be
P(d\phi)={1\over \mathcal{N}}\prod_{x} d\phi_x e^{-a^d Z^\phi_b \sum_x \phi_x (-\D+m_b^2) \phi_x))}\quad\quad 
V=a^d \sum_x [{\l_b\over 4!} \phi^4_x+{\a\over 2}
\phi^2_x]\label{33} 
\ee
where $\D$ is the discrete Laplacian and $\mathcal{N}$ the
normalization, $\l>0$. 
The expectations with respect to the measure $P(d\phi)$
are given by
$\EE(\phi_{x_1}...\phi_{2n})=\sum_\pi \prod_{(i,j)\in \pi} g(x_i,x_j)$ (Wick rule),
where the sum is over all the pairings $\pi$ of $1,..,2n$, the product is over the pairs contained in 
$\pi$ and $g(x_i,x_j)$ is the propagator, given by $g(x,y)=
{1\over L^d}\sum_{k\in [-\pi/a,\pi/a)^d} e^{i k(x-y)}
\hat g(k)$
with $\hat g(k)=(\sum_{i=1}^d (1-\cos k_i a)/a^2+m^2)^{-1}$.

By Taylor expansion then $Z=\mathcal{N}\sum_{n=0}^\io {1\over n!} E(V^n)$; the partition function is written as series in $\l_b$,
whose coefficients can be computed by the Wick rule,
and similar expressions hold for the correlations.
If the continuum $a\to 0$ 
and the infinite volume  
limit $L\to\io$ is performed, the series coincides order by with the QFT series of $\phi^4$ theory
obtained using quantum field operator. 
Each term of the series can be represented in terms of Feynman graphs which are
non connected; if we consider the expansion for $\log Z$ only connected diagrams are present.
Note that the lattice step $a$ plays the role of an ultraviolet cut-off as the momentum can reach at most values $O(a^{-1})$.

In the fermionic case ($\Phi=\psi$) then $\psi_x,\bar\psi_x$
are Grassmann variables such that $\{\psi_x,\psi_y\}= \{\bar\psi_x,\bar\psi_y\}=\{\psi_x,\bar\psi_y\}$ and $\int d\psi_x d\bar \psi_x
=0$,  $\int d\psi_x d\bar \psi_x (\bar\psi_x \psi_x)=1$; moreover
$P(d\psi)$ is a Gaussian Grassmann integration and $V$ is a sum over
monomials in the Grassmann variables. An example is 
provided by interacting Dirac fermions in $d=4$ (closely related to the Fermi model) 
\be
P(d\psi)={1\over \mathcal{N}}\prod_{x} d\bar \psi_x d\psi_x e^{-a^4 
Z_b^\psi \sum_x \psi_x (D \psi_x))}\quad\quad 
V= \l_b  \sum_x j_{\m,x} j_{\m,x}\label{22}
\ee
where $(\D\psi)_x=\sum_y \D_{x,y}\phi_y$ is the lattice Dirac derivative, which in Fourier transform
is equal to $(\hat g(k))^{-1}={1\over a} \sum_\m (\g_\m \sin p_\m a+2 \sin^2 a p_\m/2+m_b )$,$\m=0,1,2,3$
where $\g_\m$ are the gamma matrices and $j_{\m,x} =\bar\psi_{\m,x} \g_\m \psi_{\m,x}$.
$V$ is a quartic current-current interaction but other quartic intrractions can be considered, like
$(\bar\psi \psi)^2$. The fermionic expectations are given by
$\EE(\prod_i \psi_{x_i}\bar\psi_{y_i})=\sum_\pi \e_\pi\prod_{(i,j)\in \pi} g(x_i,y_j)$,
where $\e_\pi$ is the sign of the permutation.

Finally one can consider also a mixed case in which a boson and a Fermi field interact; one can consider for instance
a boson-fermion model involving $A_\m$ belonging to $\mathbb{R}^{|\L|}$
and Grassmann variables
 $\psi$,$\bar\psi$; an example is lattice QED, 
with 
\be
P(d\Phi)=P(d\psi)P(dA) \quad P(dA)={1\over \mathcal{N}}
\prod_{x} dA_x e^{-a^4 Z^A_b \sum_x A_{\m,x} (\D_{\m, \n,x} A_\n)  ))}\label{app}
\ee
where in momentum space $\D_{\m, \n,x}$ is the inverse of 
${1\over Z^A}{ 1\over |\s|^2+M^2}(\d_{\m,\n}+ {\x \bar\s_\m \s_\n\over (1-\x) |\s|^2+M^2})$
for massive vector bosons or ${ 1\over |\s|^2}(\d_{\m,\n}- (1-1/\x)  {\bar\s_\m \s_\n\over |\s|^2})$
with $\s_\m=a^{-1}(e^{i a k_\m}-1)$ for massles bosons.
The interaction $V$ is obtained replacing in the exponent of $P(d\psi)$
in each bilinear term $\bar\psi_{x\pm e_\m} \psi_x$ a term  
$\bar\psi_{x\pm e_\m}(e^{\pm i e_b  \sqrt{Z^A_b}  a A_{\m,x} }-1) \psi_x$. 
The parameters $\x$ is the gauge fixing parameter and the expectations of gauge invariant observables are $\x$-independent. 
The boson fermion model \pref{app}
is a lattice regularization of QED in the case $M=0$.

A rigorous construction of a QFT starting from \pref{1} is achieved if it is possible to
choose the bare parameters $v_b$, function of the lattice step $a$, such that the correlations admit a finite limit removing the cut-offs
$L\to\io, a\to 0$ and verify 
a set of properties known as Osterwalder-Schrader \cite{OS}
axioms. 

If the ultraviolet limit cannot be performed,
the correlations \pref{1} can be nevertheless used to define an effective QFT. The goal in this case is that the correlations of the theory
with finite cut-off can be written
as dominant cut-off independent
part, verifying continuum and relativistic symmetries, and a correction
which is of the order of the ratio of the momentum scale divided by the cut-off. 
As there is no sign of breaking of relativistic symmetries or
granularity of space time, such corrections must be very small and below the experimental resolution.
In the case of the electroweak sector, a cut-off of the order of the inverse coupling is not sufficient for modern experiments while an exponential one in the inverse coupling it is.
Another condition is that gauge invariance is valid with finite cut-off.

Expressions like \pref{1} do not compare only in QFT, but have also different physical applications.
Indeed
\pref{1} can represent the physical observables in classical Statistical Mechanics, that is averages over all the possible configurations
weighted by the Gibbs factor $e^{-\b H}$, if $\b$ is the inverse
temperature and $H$ the Hamiltonian. 
There is indeed a strict correspondence between Euclidean lattice QFT in $d$ space-time dimensions
and classical statistical mechanics models like the Ising model in $d$ space dimensions, with the Euclidean 
time plays the role of an extra space dimension. 
The Ising model 
describes the properties of magnetic materials or even lattice gases.
Such systems appear in different states or phases, characterized by the presence or absence of magnetization without  external magnetic field. 
For values of the magnetic field close to zero and temperatures close to the critical one
the thermodynamical behavior is driven by certain numbers, which can be accurately
measured and are called critical exponents. They appear to be the same in a large class
of systems, that is they are {\it universal}, a remarkable fact 
as the molecules or atoms composing such materials 
have different atomic number, interact in different way, are disposed in different lattices and so on.

It turns out that 
the system \pref{22}, describing lattice
relativistic fermions with a current interaction, describes at the same time with fixed cut-off
the Ising model in $d=2$ with a perturbation (more exactly, coupled Ising model by quartic spin interaction or veex models);
the fermionic mass in the QFT interpretation plays the role of the deviation from the critical temperature.
In the same way the model \pref{33} in a number of space-time dimensions greater than $3$ describes the Ising model 
in $d\ge3$ space dimensions, again with the mass corresponding to the deviation from the critical temperature.
The problem here consists in analyzing the long distance behavior of correlations \pref{1} 
for a fixed value of the lattice cut-off (e.g. $a=1$), in the limit of infinite volume and for vanishing or small masses,
that is close to the critical temperature.

Finally expressions like \pref{1} appear also in quantum statistical mechanics, or in condensed matter.
In this case $\L$ has to be assumed as a lattice in the space coordinate, with length $L$,
while the temporal variable has continuum values in $[0,\b]$, where $\b$ is the inverse temperature.
The equilibrium properties of non-relativistic conduction electrons in a lattice
in  a wide class of systems, including Graphene, Luttinger liquids, Weyl semimetals
or Hall insulators, whose properties are expressed in terms of interacting Dirac fermions
and whose properties are described in terms of interacting lattice Dirac fermions
expressed by \pref{22} and which can be studied by RG. More in general
phenomena like the mass generation in QCD
are strictly connected to the presence of superconductivity at low temperatures.

The fact that the same mathematical objects describe simultaneously 
QFT systems with lattice cut-off, classical statistical mechanics models and Condensed
matter systems allows to put the issue of renormalization in a more general context
and provide a deeper understanding on its meaning, thorugh the introduction of the 
Renormalization Group approach.

\section{Renormalization Group}

Let us consider the denominator of \pref{1}, the partition function $Z$.
The starting point consists in writing 
the propagator $\hat g_\Phi(k)$ as sum of propagators
\be
\hat g_\Phi(k)= \sum_{h=-\io}^N \hat g^h_\Phi(k)
\ee
with $\hat g^h_\Phi(k)=f_h(k)\hat g_\Phi(k) $ where $f_h(k)$ is a cut-off function
non vanishing for $c\g^{h-1} \le |k|\le c\g^{h+1}$ for $h\le N-1$
where $\g>1$ is an arbitrary scaling parameter and $c$ is a constant such that 
$\sum_{h=-\io}^{N-1}$ has support for $|k|\le {\pi\over 10 a}$, so that
$N\sim -\log_\g a\sim \log \L$. 
We can write
\be
\int P(d\Phi) e^{-V(\Phi)}=\int P(d\Phi^{< N})\int P(d\Phi^{N}) e^{-V(\Phi^{< N}+\Phi^N))}=
\int P(d\Phi^{< N})e^{-V^N(\Phi^{< N})}\label{44} 
\ee
where $P(d\Phi^{< N})$ has propagator  $\sum_{h=-\io}^{N-1} \hat g^h_\Phi(k)$,
$P(d\Phi^{N})$ has propagator $\hat g^{N}_\Phi(k)$
and $V^N$ is sum of monomials of any order in $\phi^{< N}$
integrated over certain kernels.
The first identity is based on the fact that 
sum of Gaussian random variables has still a gaussian distribution 
(a property holding also for Grassmann variables); in the second identity one uses that 
the Gaussian integration of an exponential is still an exponential.
In \pref{44} one integrates out the field $\Phi^N$, with momenta of order 
$O(\g^N)$, and the
the result of this integration is that $Z$ is expressed in terms of a theory 
with a lower momentum cut-off 
with a different potential.  The procedure can be iterated
integrating the fields $\Phi^{N-1},\Phi^{N-2},..$
and so on, obtaining a sequence of theories with decreasing cut-off $\g^h$ and an effective potentials $V^h$. 

During such procedure some of the terms in $V^h$ tend to increase and others to decrease.
A simple criterion to understand which one is increasing or decreasing 
is based on scale invariance of the propagator, that is the fact that
\be
g^{\le h}_\Phi(x)\sim \g^{(h-N)(d-\a)}g^{\le N}_\Phi(\g^{(h-N)} x)
\ee
where $\a=2$ for boson and $\a=1$ for fermions.
The field with cut-off at scale $h$ 
$\Phi^{\le h}(x)$ is essentially distributed as $\g^{(h-N)(d-\a)/2}$
$\Phi^{\le N}(2^{(h-N)} x)$, up to a rescaling and a prefactor.
The monomials appearing in $V^h$ of the form $\int dx O^{\le h}_{n_\phi, n_\psi}$, if $O_{n_\phi, n_\psi}^{\le h}$
is a local monomial with $n_\phi$ bosonic and $n_\psi$ fermionic fields, 
behaves essentially as
\be
\g^{-(h-N) D_{n_\phi,  n_\psi} }\int dx O_{n_\phi, n_\psi}^{\le N} \quad
D_{n_\phi,n_\psi}=d-{(d-2)\over 2} n_\phi-{(d-1)\over 2} n_\psi
\ee
Therefore after the integration of the field at scales $N,N-1,..,h$ one obtains an expression similar to the initial one but where each monomial in the potential is multipled by a factor $\g^{-(h-N) D_{n_\phi,n_\psi}}$.
The terms with $D_{n_\phi,n_\psi}>0$ are increased integrating the fields from scale $N$ to scale $h$;
they are called {\it relevant}. The terms $D_{n_\phi,n_\psi}<0$ are decreased and are called irrelevant.
The terms with $D_{n_\phi,n_\psi}=0$ are unchanged by this scaling argument and are called marginal.

The integration of the field $\Phi^h$ must be done separating the irrelevant terms from the others. It is also convenient to write
the marginal or relevant terms as local ones, that is with all the fields computed in the same point; this is always possible at the price of extracting
enough irrelevant terms. In addition, it is also convenient to include the quadratic local terms in the gaussian, modifying
the single scale propagator.
One obtains therefore an expression of the form
\be
\int P(d\Phi) e^{V(\Phi)}=\int P(d\Phi^{\le h}) e^{-\mathcal{L} V(\Phi^{\le h})-\mathcal{R} V(\Phi^{\le h})  }\label{44a} 
\ee
where $P(d\Phi^{\le h})$ is gaussian integration with a propagator with 
a cut-off selecting momenta $|k|\le \g^h$, with mass $m^\Phi_h$ and wave function renormalization $Z_h^\Phi$,
$\mathcal{L}  V(\Phi^{\le h})$ is sum of monomials corresponding to the marginal or relevant local terms (excluding the ones contributing to $m_h$ or 
$Z^\Phi_h$
) and with coupling $v_h$
depending on the scale $h$, and 
$\mathcal{R}  V(\Phi^{(< h)})$ is sum of irrelevant monomials and are typically non local and integrated over kernels.
The couplings $v_{h}$ appearing in $\mathcal{L}  V_h$ are called running coupling constants (rcc). The correlations can be 
written as derivative of generating functions $\log \int P(d\Phi) e^{-V(\Phi)-B(\Phi,J)}$, where $B$ is a source term,
that is a sum of monomials in the external fields, and it could be analyzed by a similar analysis.  
From the physical point of view, \pref{44a} says that the constants of a theory are not constant at all;
they are different at different energy scale.
The physical quantities, and the kernels of $\mathcal{L}  V^h$, can be written as power series of the 
running coupling constants, which is often represented in term of trees \cite{G}. 

We consider $h=0$ the energy scale at which
the dressed constants are fixed. The ultraviolet or high energy (small distances) problem consists in investigating
the behavior at positive scales; in particular
in order to construct a QFT we have to prove that it is possible to choose the bare constants, that is the constants at the fundamental energy scale $N$ (extremely small distances)
$v_N=v_b$,
so that the dressed constants, that is the ones at scale $h=0$, have a fixed value $v_0=v_d$. The infrared problem consists in studying 
the behavior for negative scales (large distances); this is relevant for massless QFT models and statistical physics or condensed matter,
in which only negative scales are present. In this case $h=0$ represent the microscopic scale and $h=-\infty$ the macroscopic scale of matter.

In a QFT
all the irrelevant terms present and finite at the ultraviolet scale $N$
disappear or become very small at the observable scale $0$. The theory at the measurable scales depend only on the marginal or relevant terms
present in the bare action, and is insensitive to the details of the ultraviolet scale, that is to the details of the (presumed) more fundamental theory describing extreme energies.
On the other hand
such a feature
in the application to Statistical mechanics, like in the theory of critical phenomena,
is at the basis of the universal behavior observed close to phase transitions, as the theory is largely independent from the 
details present at scale $h=0$. 

Let us consider first a non-renormalizable theory as a fermionic theory like
\pref{22} in $d=4$ where 
the quartic interaction is irrelevant in the RG sense. The coupling 
decrease as $\l_b 2^{2(h-N)}$ so if we want to fix the coupling at scale $0$ to a finite
value $\l_d$  (for instance the Fermi constant in the application to weak forces), then the bare constant has to be $O(\l_d 2^{2N})$, hence
one cannot take the limit $N\to \io$ to have a finite bare constant; one needs
$\l_d 2^{2N}=O(1)$, that is the cut-off is of order of the inverse of the coupling. 
A possibility for reaching higher cut-off is that there is a "non-trivial fixed point", that is the effective coupling stop increasing but it reaches a finite value so that the ultraviolet cut-off could be removed \cite{GK0}.
On the other hand, integrating the negative
scales the effective interaction becomes smaller and smaller; one expects that the large distance behavior is the same as the 
non interacting case. This explain in particular the fact that interactions do not modify the properties of materials like graphene. 

Let us consider now a renormalizable theory. In this case the interaction have zero dimension, that is they are marginal.
The outcome of RG is that the physical observables
are written as a series in the
rcc. They can be again expressed in terms of Feynman graphs but now they are bounded
even removing the ultraviolet cut-off $a\to 0$ and the infrared one $L\to\io$,
in the massless case and one can typically prove bounds $O(n! |v_h|^n)$.
There are no divergences, while in the expansion in the bare coupling in $V^N$
the coefficients were divergent when cut-offs are removed. Of course the rcc are functions of the bare couplings,
and if one re-expand them divergences are re-encountered again. 

The consistence of the method relies on the fact
that the rcc remain small and this depend on the system one is considering.
The fact that the effective couplings are marginal only means that
they do not increase or decrease exponentially, but nevertheless they are not necessarily constants.
Indeed the marginal coupling obey to recursive equations called beta function; in the case of positive beta function they decrease
increasing the scale (marginally irrelevant), if negative
increase
increasing the scale (marginally relevant) and if
asymptotically vanishing $O(\g^h)$ it remain close to its initial value.
The basic case of QED is marginally irrelevant and
\be
e^2_h\sim {e_d^2\over 1-\log(\g^h)  {e_d^2\over 6\pi^2}
}\label{G}
\ee
where 
$e^2_N=e^2_d\sim {e_d^2\over 1-\log(\g^N)  {e_d^2\over 6\pi^2}}$.
In order to avoid that $e_d$ becomes too large one needs a maximal allowed cut-off $\L$ or $\g^N$ which is $O(e^{e_d^{-2}})$, that is exponentially high in 
the inverse charge, much higher than in the non-renormalziablr case were it was of the order of the inverse coupling.
One expects also that the other marginal couplings are small 
or vanishing on this regime, as consequence of Gauge symmetry. Therefore one can consider QED as an effective theory, 
with a cut-off which is 
exponentially high in the inverse of the bare coupling, which is much higher than any energy reachable in experiments.
This is consistent with the fact that no sign of granularity of space-time
or other effects due to cut-offs are seen experimentally.
In the infrared regime, the interaction becomes weaker and weaker so that the particles behaves as free at large distances.
A similar behavior is found also in the $\phi^4$ in $d=4$, that is the ultraviolet cut-off cannot be removed but the infinite volume massless limit can be studied, and the fact that the effective coupling decreased implies the validity of mean field theory for Ising model in $d\ge 4$. Also in the Standard electroweak model
the coupling is marginally irrelevant if the anomaly cancel out,
hence the theory could be constructed with exponentially high cut-off.

The ultraviolet cut-off 
cut-off can be removed in the case
of marginally relevant coupling and the theory can be perturbatively constructed in principle removing cut-off (but preserving the finite volume). 
In this case one says that there is asymptotic freedom
and this is what happens in QCD, the sector of strong forces of the Standard Model, in which
in principle one could construct the theory verifying the axioms.
The increasing of the effective coupling means that the behavior of correlations
is strongly modified at large distances and 
a spontaneous generation of mass occurs, while particles are essentially non interacting at small distances. 
Another case in which the cut-offs can be removed is when
the Beta function is vanishing, so that
the theory remain weakly interacting both for large or small energies. In contrast with asymptotically free models,
the physical properties of the theory 
are modified with respect to the non-interacting case. Typically the scaling dimensions
of the fields are different and continuos function of the coupling, as happens in the Fermi theory \pref{22} with $d=2$.

In conclusion
RG shows that
there is a sequence of effective models
living at different energy scales. Instead of considering the bare and dressed theory, there is sequence of theories interpolating between the two.
The parameters appearing in the theory, like the mass or the coupling, appear to be dependent
on the energy scale. It can even happen that the effective theories
at different scales can be different not only for the value of the parameters but in a more drastic way
and radically different description can emerge iterating the RG.

\section{Rigorous results}

Renormalization, especially in its Renormalization Group version, has a rigorous formulation,
see e.g. \cite{1a}- \cite{5}, and in several cases a mathematical non-perturbative construction of several models has been reached.  
In fermionic models in $d=2$ with a quartic interaction
the complete program of renormalization has been performed.
Two main models in this class are the Thirring model \cite{Th}
and the Gross-Neveu model \cite{GN}
and can be used as a test bench 
for higher dimensional QFT. 
In both cases
the superficial degree of divergence is $D=2-n_\psi/2$ hence
they are renormalizable models and the quartic coupling has zero dimensions. In one case the interaction
has the form $\l_b\int dx j_\m j_\m$ while in the second
the fermions have a color index $i=1,..N_c$, with $N_c\ge 2$ and the interaction has the form ${\l_b\over N_c} \int dx  (\bar\psi_x\psi_x)^2$.
The starting point is consider the theory with a cut-off, which can be a lattice one, as in \pref{22}, or a momentum one.
As an outcome of the RG analysis it is proved that the correlations can be written as series 
in the running coupling constant $\l_h$ which are analytic around the origin. 
The expansion in Feynman graphs has to be avoided; the graphs are indeed finite uniformly in the cut-offs
but their number is too large so that one gets in this way only an $n!$-bound 
which is not sufficient to prove the absolute convergence. It is necessary
to take
into account compensations between the graphs contributing to a certain order, which are due to relative sign between them ultimately following from Pauli principle.
This can be implemented using cluster expansions formulas 
for truncated expectations \cite {Br}, \cite{Le}, writing them as sum over trees of propagators times determinants, which
can be bounded by Gram inequality $|\det A|\le C^n$
without producing factorials; the is one get a bound at order $n$ for the correlations $C^n |\l_h|^n$ which implies analyticity in the rcc $\l_h$. 
 If one re-expand the effective coupling $\l_h$ one gets a series
presenting $O(\log\L^n)$ at order $n$, so recovering the ultraviolet divergence of the perturbative expansion. 
The construction of the models requires that the bare couplings can be
chosen so the the running coupling constant do not exit from the convergence radius.

In the Gross-Neveu
the effective coupling is marginally relevant and it is possible to choose, for $\l_d>0$ and small enough
\be
\l_b= {\l_d\over 1+
{2\over \pi}\l_d^2(N-1)\log \L(1+O(\l_d))}\quad \quad Z_b=1+O(\l_b)\quad m=O(1)
\ee
so that the Schwinger functions of the model are finite in the limit $\L\to\io$ and in the infinite volume limit \cite{GK}, \cite{FMRT}
and the axions are verified. The construction requires a large fermion mass and the correlations are Borel summable as function of $\l_d$. As in the case of QCD, if the infrared cut-off is removed in the massless case one expects a fermionic mass generation.

In  
the Thirring model the coupling is marginaly marginal.
By choosing 
\be
{\l_d\over (Z^\psi_b)^2}=\l_b+O(\l_d^2)\quad Z_b= \L^{-\h}\quad m_b=m \L^{\h_m} \quad\h=\l_d^2+O(\l_d^2),\quad \h_m=O(\l_m)
\ee
it was proved in \cite{BGM1} that the Schwinger functions in the limit $N\to\io$ exists and are analytic
for $\l_b$ small enough, uniformly in the fermionic mass and in the volume. The correlations are analytic 
in $\l_d$ in a disk around $\l_d=0$.
The bare coupling and wave function renormalization are not independent but verify
a non trivial relation, analogue to \pref{imp} in QED; divergences in the effective quartic coupling must be exactly compensated by the divergences in the wave function normalization. 
In the RG analysis the quartic coupling can be written as $\l_h Z_h^2$,
and one has to show that $\l_h=\l_b+O(\l_b^2)$, a fact following from complicate cancellations in the Beta function. They cannot be established
by direct computation but follows by a combination of Ward Identities
and Schwinger-Dyson equation at each Renormalization Group iteration. 
One can remove the ultraviolet cut-off and infrared cut-off and the massless limit can be performed. The asymptotic behavior of correlations is modified with respect to the non.interacting case by the 
presence of critical exponents, and the non-renormalzation of the 
anomaly can be proved.  With a fixed ultraviolet cut-off $\L=1$
the model defines a wide
class of universality including Condensed matter systems like Luttinger liquids or statistical mechanics model like non solvable interacting Ising, Vertex or Dimer models \cite{BGM2}.

In the case of bosonic models, like $\phi^4$ models in $d=4$,
the coupling is marginally irrelevant and it has 
 been proved in
\cite{GK1}, \cite{GK2} that the infinite volume 
and massless limit can be constructed rigorously; in order to describe a theory with vanishing dressed mass one needs to choose properly
the bare one. Again one has to avoid the Feynman graph expansion by using
cluster expansion formulas dividing regions 
in which the field is large, where
one uses that $e^{-\l\sum \phi^4}$ is small, from  
regions in which is small, in which one can expand. The result has important implication in statistical mechanics, where it implies the 
validity of mean field theory in the Ising model at $d=4$, up to logarithmic corrections. Deformations of $\phi^4$ theory, in which the boson propagator is $p^{-1/2(3+\e)}$ have been considerate in $d=3$ 
\cite{BMS},  where the existence of a bosonic fixed point is proved,
somewhat related to the one found with the Wilson $\e$-expansion. A similar construction of a non trivial fixed point can be done also for fermions \cite{GMR}.

In particle physics theory and in particular in the Standard Model, where the notion of renormalization was originally introduced, one should distinguish between the electroweak sector and the strong one.
If one considers the second one the theory is expected to be marginally relevant (asymptotically free in the ultraviolet) and in principle one could remove the ultraviolet cut-off.
In the case of pure Gauge fields, that is considering the Yang-Mills theory
in $d=4$,
without particle fields the removal of the ultraviolet cut-off has been studied in 
\cite{Ba1},\cite{Ba2} with an infrared volume cut-off by using a probablistic approach taking into account Gauge invariance. 
Removing the infrared cut-off proving the mass generation is a major open problem.
Similar methods can be applied to the construction of QED in $d=3$ \cite{Ba3}, corresponding to a case in which the interaction is relevant so that the ultraviolet cut-off can be removed assuming a large fermionic mass. Considering the full Standard model the theory can be probably constructed only as an effective one with a finite cut-off, as the electroweak sector is marginally irrelevant.  

At lower energies the Standard electroweak sector reduces essentially
to the Fermi theory, which is a fermionic theory of the form \pref{22},
combined with QED.
For such systems the existence of the infinite volume limit for massless fermions
and massive photons, with an ultraviolet cut-off can be proved, that is in (2.4) $|P_n|\le C^n\L^n $ so that analyticity follows for $\L^2 C G<1$, that is up to a cut-off of the order of $G^{-2}$.
In addition, one can establish the non-renormalization of the chiral anomaly \cite{GMP1}, a crucial property for the construction of chiral Gauge theory.  Fermionic models with quartic interaction with cut-off $\L=1$ are used to describe the physical properties of Weyl semimetals in $d=4$ or graphene
in $d=3$ \cite{GMP2}, and the universality properties of the anomaly
are strictly related to the universality properties of transport coefficients. 
The construction of 
QED and the electroweak Standard Model 
as effective theory up to an exponentially high cut-off is also an open problem.

\end{document}